\documentclass[preprint,12pt]{elsarticle}

\usepackage{amssymb}
\usepackage{amsmath}

\usepackage{booktabs}
\usepackage{pifont}
\usepackage[table]{xcolor}   
\usepackage{multirow}
\usepackage{adjustbox}
\usepackage{hyperref}

\journal{Signal Processing}

\begin{document}

\begin{frontmatter}



\title{Towards Open World Sound Event Detection}


\author[label1]{Pham Hoang Hai\fnref{equal}} 
\ead{21020015@vnu.edu.vn}
\author[label2]{Le Trong Minh\fnref{equal}\corref{cor1}} 
\ead{21020355@vnu.edu.vn}
\author[label2]{Le Hoang Son\corref{cor1}} 
\ead{sonlh@vnu.edu.vn}

\fntext[equal]{These authors contributed equally to this work.}
\cortext[cor1]{Corresponding author}

\affiliation[label1]{organization={VNU University of Engineering and Technology},
            city={Hanoi},
            postcode={100000}, 
            country={Vietnam}}
\affiliation[label2]{organization={Artificial Intelligence Research Center, VNU Information Technology Institute},
            city={Hanoi},
            postcode={100000}, 
            country={Vietnam}}

\begin{abstract}
Sound Event Detection (SED) plays a vital role in audio understanding, with applications in surveillance, smart cities, healthcare, and multimedia indexing. However, conventional SED systems operate under a closed-world assumption, limiting their effectiveness in real-world environments where novel acoustic events frequently emerge. Inspired by the success of open-world learning in computer vision, we introduce the Open-World Sound Event Detection (OW-SED) paradigm, where models must detect known events, identify unseen ones, and incrementally learn from them. To tackle the unique challenges of OW-SED, such as overlapping and ambiguous events, we propose a 1D Deformable architecture that leverages deformable attention to adaptively focus on salient temporal regions. Furthermore, we design a novel Open-\underline{W}orld Def\underline{O}rmable S\underline{O}und Event Detection \underline{T}ransformer (WOOT) framework incorporating feature disentanglement, separating class-specific and class-agnostic representations, and a one-to-many matching strategy with a diversity loss to enhance representation diversity. Experimental results demonstrate that our method achieves marginally superior performance compared to existing leading techniques in closed-world settings and significantly improves over existing baselines in open-world scenarios.
\end{abstract}




\end{frontmatter}




\section{Introduction}
Sound Event Detection (SED) is a core task in audio understanding, with applications spanning  surveillance~\cite{crocco2016audio}, smart cities~\cite{salamon2018sonyc}, medical monitoring~\cite{phuong2013sound}, and multimedia indexing~\cite{hershey2017cnn}. Traditional SED systems are typically developed under a closed-world assumption: all sound event classes that may appear during inference are known in advance and included in the training set. While this setup has driven progress in benchmark-driven ~\cite{adavanne2017sound, sedt, cnn, li2020sound, crnn}, it falls short in real-world scenarios, where systems are often deployed in dynamic and unpredictable environments and must handle previously unseen events.

To address similar challenges in the visual domain, the Open-World Object Detection (OWOD) paradigm was introduced \cite{joseph2021towards, prob, cat, ow-detr}. OWOD systems aim to detect known categories while identifying previously unseen objects, flagging them for annotation, and incrementally learning new categories without catastrophic forgetting. This open-world formulation has significantly advanced the robustness and adaptability of object detectors in real-world environments.

\begin{figure} [h]
    \centering
    \includegraphics[width=0.9\linewidth]{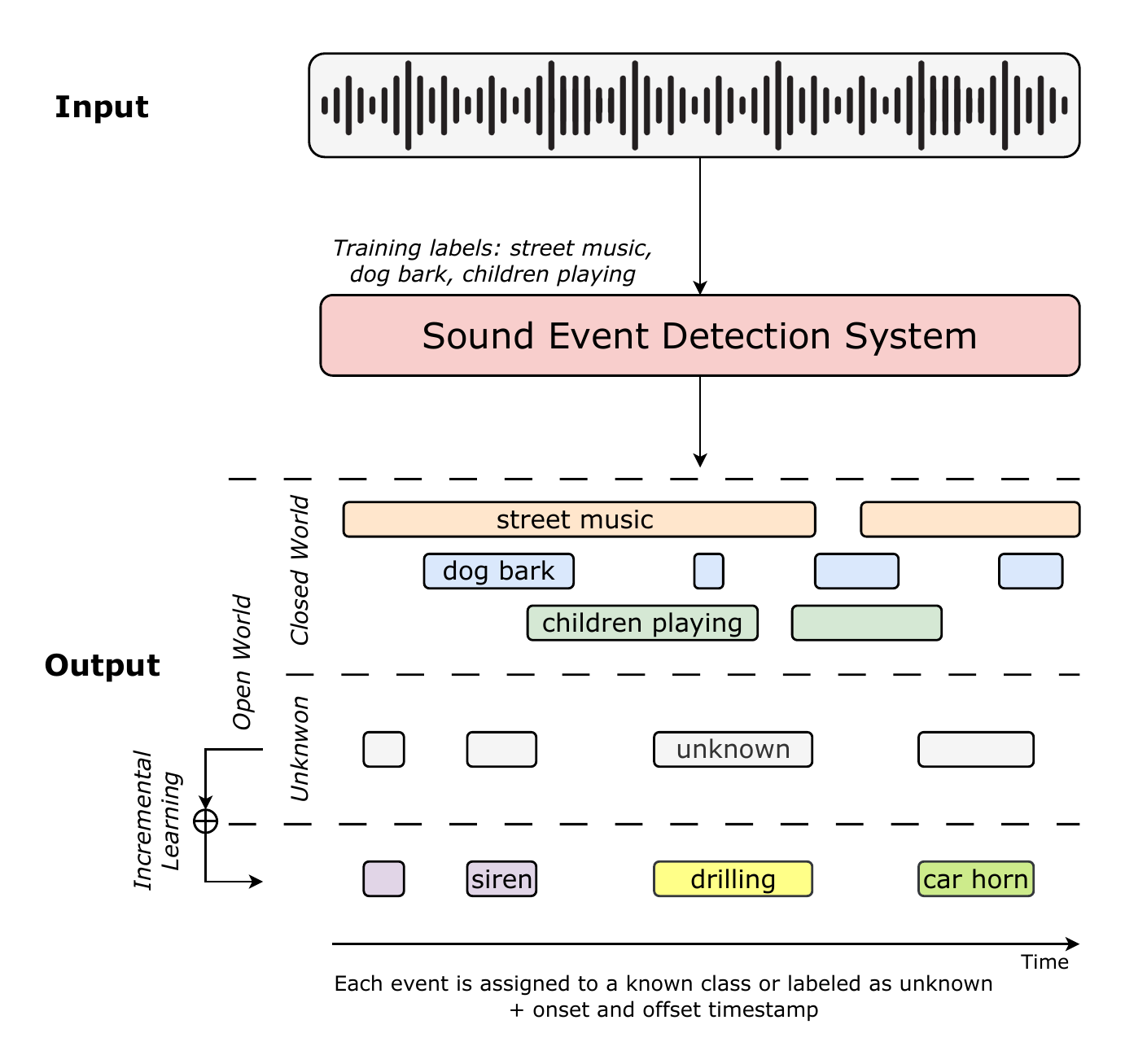}
    \caption{Introduction to the Open-World Sound Event Detection (OW-SED) task}
    \label{fig:owsed}
\end{figure}

Inspired by this paradigm, we propose the first formulation of open-world learning for Sound Event Detection, termed Open-World Sound Event Detection (OW-SED). As illustrated in Figure~\ref{fig:owsed}, OW-SED extends conventional Sound Event Detection by requiring a model not only to identify sound events from a set of known training classes (e.g., street music, dog bark, children playing) but also to recognize the presence of unknown acoustic events during inference. These unknown events can then be labeled by a human oracle and incrementally integrated into the model, thereby enabling continual learning of new sound classes (e.g., siren, drilling, car horn). Compared to vision, sound events are typically temporally overlapping, ambiguous, and context-dependent, posing unique challenges for open-world modeling. Our work introduces OW-SED as a new and necessary direction for robust, adaptive, and realistic audio understanding systems.

In this paper, we first propose 1D Deformable architecture for sound event detection. In SED, consecutive time segments often exhibit high similarity due to slow-varying background sounds and redundant acoustic patterns. Moreover, sound events frequently lack clearly defined temporal boundaries and may overlap with one another, making precise temporal localization especially challenging. These characteristics demand that SED models be highly sensitive to subtle, local variations in the temporal domain. However, standard Transformer architectures, used in previous work~\cite{sedt}, treat all positions in the sequence equally, making them less sensitive to subtle local changes. To address this challenge, we adopt deformable attention~\cite{d-detr}, which focuses on a limited number of relevant positions surrounding a reference point within the input sequence. Both the sampling offsets and corresponding attention weights are learned and dynamically tuned based on the input, allowing the model to adaptively focus on informative regions while preserving locality awareness. 

In addition, we introduce a novel framework named \textbf{Open-World DefOrmable SOund Event Detection Transformer (WOOT)}, which builds upon our proposed 1D Deformable architecture and prior work on open-world object detection~\cite{prob}. Our framework incorporates two additional key improvements to better address the challenges of the Open-World Sound Event Detection domain. First, we propose to disentangle the feature representation of each detected event into a class-specific component and a class-agnostic component. This separation encourages the model to better generalize to unseen classes by isolating information that is invariant across sound event categories. Second, we introduce a two-stage training process. In a standard detection training pipeline, each ground-truth event is matched to only one query via Hungarian matching. This one-to-one constraint ignores other queries that are not matched, even if they make reasonable predictions (e.g., queries that predict the correct class and whose predicted segments are fully contained within the corresponding ground-truth interval). To overcome this limitation, we adopt a one-to-many matching strategy in the first stage, allowing multiple queries to be matched to the same ground-truth event. However, this approach can lead to redundant optimization, where many queries are trained to represent the same ground-truth event. As a result, the model's capacity to capture diverse patterns, particularly those useful for unknown class detection, becomes limited. To mitigate this, we introduce a diversity loss in the second stage of training, which explicitly encourages the features of different queries to be dissimilar. This promotes a more diverse set of learned representations, boosting the model’s effectiveness in detecting and representing novel sound events.

Experiments on the URBAN-SED~\cite{urban-sed} and DESED~\cite{desed} dataset show that our 1D Deformable model performs competitively with strong existing methods in the closed-world setting, while our open-world framework achieves superior performance in the open-world setting compared to prior baselines.

\textbf{In summary, the core contributions of our paper are listed below}:
\begin{itemize}
\item We introduce the first formulation of Open-World Sound Event Detection (OW-SED), extending the open-world learning paradigm from computer vision to audio understanding, where models must detect known sound events while identifying and learning from previously unseen acoustic events.
\item We present a novel 1D Deformable architecture designed for sound event detection that uses deformable attention to adaptively focus on informative temporal regions, addressing the challenges of temporally overlapping, ambiguous, and context-dependent sound events.
\item We develop a novel Open-World Deformable Sound Event Detection Transformer (\textbf{WOOT}) framework that incorporates two key improvements: feature disentanglement that separates event representations into class-specific and class-agnostic components for better generalization to unseen classes, and a training process combining one-to-many matching with diversity loss to promote diverse learned representations while avoiding redundant optimization.
\item We demonstrate through experiments on the URBAN-SED~\cite{urban-sed} and DESED~\cite{desed} datasets that our approach outperforms existing techniques, achieving state-of-the-art results.
\end{itemize}

\section{Related Works}
\subsection{Sound Event Detections}
In recent years, there has been growing attention on Sound Event Detection, which focuses on identifying sound event categories and localizing their temporal onsets and offsets within an audio signal. This surge in interest is largely driven by its wide range of applications and the advancements in deep learning technologies. Early research in SED primarily focused on leveraging traditional machine learning models with hand-crafted features. Approaches utilizing Gaussian Mixture Models and Hidden Markov Models, often applied to Mel-frequency cepstral coefficients or other low-level descriptors, were the standard practice for modeling acoustic events \cite{7096611, 7100934}. However, these approaches relied on simplistic statistical models with limited capacity to model hierarchical or non-linear patterns in acoustic signals, making them inadequate for capturing the spectral and temporal complexities of real-world soundscapes involving overlapping or context-dependent events. 

The advent of deep learning introduced a fundamental shift in methodology. Convolutional Neural Networks (CNNs) were first employed to learn structured patterns in the time–frequency domain from log-mel spectrograms and other time–frequency representations \cite{7324337}. This was soon followed by the incorporating of Recurrent Neural Networks (RNNs), particularly Long Short-Term Memory (LSTM) units, leading to the widely adopted Convolutional Recurrent Neural Network (CRNN) architecture \cite{crnn}. CRNNs became the dominant model for SED tasks in the DCASE challenge series, consistently outperforming previous approaches. Building upon this, Nam et al. \cite{fdy-crnn} introduce the Frequency Dynamic Convolutional Recurrent Neural Network (FDY-CRNN), a novel convolution module to improve the model’s capability in capturing diverse frequency patterns, resulting in improved performance on domestic
environment sound event detection (DESED) dataset. 

More recent work has explored attention-based and transformer-based architectures, which enable models to focus on relevant parts of the input sequence and handle long-range dependencies without recurrence. For instance, the Audio Spectrogram Transformer for SED (AST-SED) \cite{ast-sed} combines a frequency-wise transformer encoder with a temporal Bi-GRU decoder to restore temporal resolution. In parallel, Conformer-based architectures \cite{conformer}, which combine convolutional modules with self-attention, have shown great promise for polyphonic SED. In particular, Barahona et al. \cite{conformer-sed} proposed a Conformer-based system enhanced with Frequency Dynamic Convolutions and BEATs audio embeddings, which significantly improved Polyphonic Sound Detection Score (PSDS). Besides, the Sound Event Detection Transformer (SEDT) \cite{sedt} has emerged as a novel paradigm that reconceptualizes SED as a set prediction problem, drawing inspiration from the Detection Transformer (DETR) \cite{detr} framework originally developed for computer vision. This end-to-end architecture introduces a one-dimensional variant of DETR specifically adapted for temporal sequences, featuring a one-to-many bipartite matching strategy and an audio query branch to better capture category-specific information and improve classification performance. More recently, Yin et al.~\cite{YIN2025109691} introduced a multi-granularity acoustic information fusion approach that leverages an interactive dual-conformer module to capture complementary fine- and coarse-scale temporal features, further advancing detection performance in complex acoustic scenes.


Despite these advancements, existing SED systems predominantly operate under the closed world assumption, where the set of target sound classes is fixed and known during training. This constraint limits the generalization capability of current models when deployed in dynamic and evolving acoustic scenes.

\subsection{Open-set and Open-Vocabulary Sound Event Detection}
Some recent works have explored open-set classification and open-vocabulary settings that relax the closed-world label assumption in SED, but they do so in different ways. You et al.\cite{osse} study about open-set sound event classification primarily focuses on unknown rejection, where the model decides whether an input belongs to one of the known classes or should be rejected as unknown, without considering temporal boundary localization. In contrast, open-vocabulary SED methods aim to detect and label events beyond the training ontology through audio–text models, in which the detector is conditioned on free-form textual or audio queries and localizes events that match the query semantics, as in Detect Any Sound~\cite{detectanysound} and FlexSED~\cite{flexsed}. \textbf{OW-SED} differs from both settings in what is assumed at deployment and what is required over time. It operates in a prompt-free manner and must detect and localize previously unseen sound events as unknown during inference, while also enabling these unknowns to be incorporated as new known classes in later learning phases. 

Additionally, several recent works have investigated class-incremental learning for SED. Representative examples include UCIL \cite{ucil} and the framework of Pandey et al. \cite{pandey2024classincrementallearningsoundevent}. These methods assume that, at each learning phase, audio segments from newly introduced sound classes are already labeled and assigned to known class identities, and the main objective is to update the detector while mitigating catastrophic forgetting. However, Open-World SED is broader than this setting. It addresses not only how to learn from newly provided classes, but also how such classes emerge in the first place. At deployment, novel sound events are neither labeled nor specified in advance and must first be detected and localized as unknown before they can be incorporated into later learning stages. This formulation follows the Open World Learning (OWL) perspective, a well-established task setting (discussed in more detail in \ref{sec:owl}), which couples known-class recognition, unknown identification, and progressive learning of novel classes over time. In this way, Open-World SED naturally combines Open-set detection with class-incremental SED, resulting in a more general and realistic learning paradigm for sound event detection.

\subsection{Open World Learning} \label{sec:owl}
Open World Learning is a learning paradigm designed to address the limitations of conventional closed-world systems by enabling models to recognize known classes while simultaneously identifying, rejecting, and eventually learning novel, previously unseen categories during deployment. One of the most actively studied domains in OWL is Open World Object Detection (OWOD). Joseph et al. \cite{joseph2021towards} formalized the OWOD problem and introduced the Open World Object Detector (ORE), which combines an energy-based unknown object classifier with contrastive clustering to discover and integrate novel categories over time. Subsequent studies have focused on transformer-based architectures. OW-DETR (Open-world Detection Transformer) \cite{ow-detr}, introduced by Gupta et al., is a transformer-based framework that addresses OWOD challenges through attention-driven pseudo-label generation, a classifier for distinguishing novel categories, and an objectness evaluation module. This method utilizes a pseudo-labeling scheme where object queries with strong attention scores that do not align with any known class annotations are selected as pseudo-unknowns. Similarly, Open World DETR \cite{openworlddetrtransformer} builds on Deformable DETR with a two-stage training method, incorporating a class-agnostic binary classification head and a multi-view self-labeling mechanism that generates pseudo ground truths for unknown classes by combining a pre-trained binary classifier with selective search. In contrast, PROB \cite{prob} introduces a novel probabilistic framework for objectness estimation that does not use pseudo-labeling. It separates objectness prediction from classification using a density-based objectness head, enabling the detection of unknown objects without relying on negative examples or explicit background-unknown separation. Another distinct approach, OW-RCNN \cite{ow-rcnn} adapts the Faster R-CNN architecture to the open-world setting, introducing a class-agnostic Region Proposal Network (RPN) and a discriminative unknown-aware classifier, and supports continual learning through a decoupled representation and classification head.

In addition to OWOD, the principles of Open World Learning have been explored across various domains. In image classification, Bendale \cite{openmax} introduced the OpenMax framework, which estimates whether a test sample belongs to an unknown category by modeling activations in the penultimate layer using class-wise Weibull distributions, enabling open set recognition without retraining. Shu et al. \cite{doc} proposed DOC (Deep Open Classification), which utilizes one-vs-rest sigmoid output layers with a confidence threshold to separate known and unknown classes, allowing neural networks to reject unseen inputs during inference. In medical imaging, Zamzmi et al. \cite{medical} proposed an open-world active learning framework for echocardiography view classification that labels known views, clusters unknown ones for expert review, and incrementally updates the model to handle new imaging views. While open-world learning has been mostly studied in vision, related work in audio has also started to surface in the form of anomalous sound detection. For example, Zheng et al. \cite{ZHENG2026110218} propose STWWgram-ODCBAM, which fuses multimodal features and attention mechanisms to detect abnormal acoustic events in industrial settings.

Collectively, these works demonstrate the broad applicability and importance of OWL across domains where static closed-world assumptions are insufficient for real-world deployment. In this research, we take the first step toward exploring Open World Learning for the task of Sound Event Detection. We formally define the OW-SED problem, propose a baseline framework inspired by open-world object detection principles, and establish a benchmark for evaluation. This work marks the first attempt to extend OWL into the audio domain, paving the way for developing adaptive and robust SED systems capable of handling novel sound events in dynamic acoustic environments.

\section{Problem Formulation}
At time $t$, the known set of sound event classes is defined as $\mathcal{K}_t = \{1, 2, \dots, C\}$ where $C$ denotes the current total of identified and learned categories. Let $\mathcal{D}_t = \{A_t, Y_t\}$ be the dataset at time $t$, where $A_t = \{a_1, \dots, a_N\}$ is a collection of $N$ audio clips and $Y_t = \{y_1, \dots, y_N\}$ denotes the corresponding event annotations. Each label $y_i = \{v_1, \dots, v_K\}$ is a set of $K$ sound events in clip $a_i$, where each event $v_k = [l_k, s_k, e_k]$ consists of the class label $l_k \in \mathcal{K}_t$ and its onset $s_k$ and offset $e_k$ timestamps.

Let $\mathcal{U} = \{C+1, C+2, \dots\}$ represent the set of unknown classes that might occur during inference but are not included in the current training data. In the OW-SED setting, the model $f_t$ is trained at each time step $t$ to identify sound events belonging to the known classes $\mathcal{K}_t$, while also identifying novel acoustic events by assigning them a special unknown class label (denoted as class 0). These unknown predictions are then reviewed by a human oracle, who labels a subset $\mathcal{U}_t \subset \mathcal{U}$ with $n$ new class identities and supplies a corresponding set of labeled training instances.

After this update, the set of known classes becomes $\mathcal{K}_{t+1} = \mathcal{K}_t \cup \{C+1, \dots, C+n\}$. To simulate realistic constraints such as limited memory, privacy, and computational cost, only a small portion of past data can be retained. As a result, the model must be incrementally updated from $f_t$ to $f_{t+1}$ without retraining from scratch. The updated model $f_{t+1}$ should be able to detect all classes in $\mathcal{K}_{t+1}$ while maintaining performance on previously learned classes, thus avoiding catastrophic forgetting.

This process continues over time as new unknown events are discovered, labeled, and incrementally learned. The goal of OW-SED is to support this continual cycle of open-world learning while maintaining robust detection and generalization across both seen and unseen acoustic events.

\section{Methods}
\subsection{1D Deformable DETR}
\begin{figure} [t!]
    \centering
    \includegraphics[width=1\linewidth]{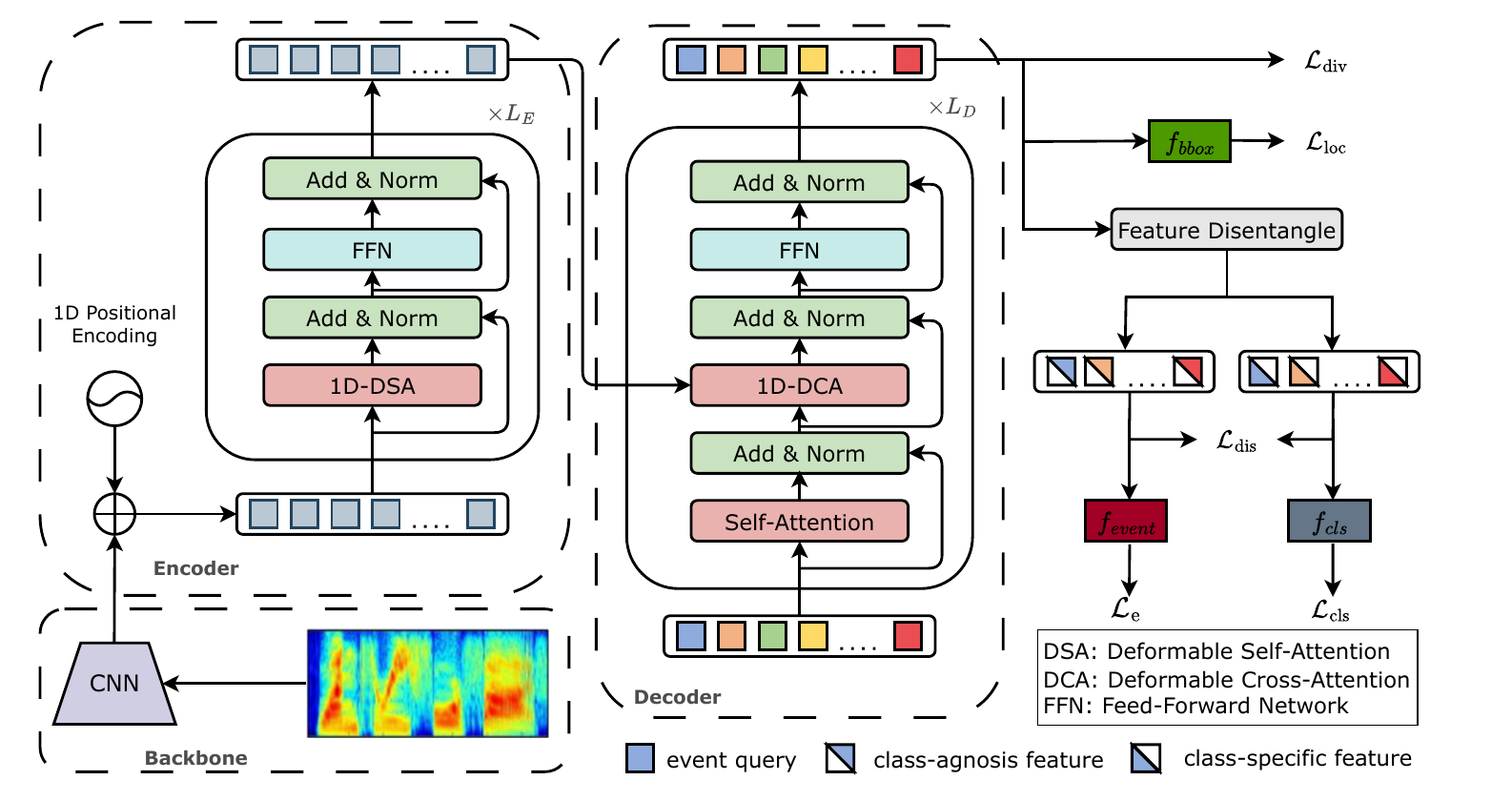}
    \caption{\textbf{Illustration of the WOOT model architecture}. The proposed WOOT is built upon a 1D Deformable Transformer backbone specifically tailored for sound event detection. It introduces a transformer encoder with \textit{1D Deformable Self-Attention} (1D-DSA) to enable efficient temporal modeling, a decoder with \textit{1D Deformable Cross-Attention} (1D-DCA) to progressively refine event representations, and a specialized prediction head comprising a regression branch ($f_{bbox}$) for estimating event timestamps, a classification branch ($f_{cls}$) for computing class probabilities, and an eventness branch ($f_{event}$) for evaluating eventness scores. On top of this backbone, WOOT further introduces two key components to address the challenges of open-world sound event detection: (i) \textit{Disentangled feature representation}: each detected event is decomposed into a class-specific and a class-agnostic component, enhancing generalization to unseen sound classes. (ii) \textit{Two-stage training strategy}: the first stage adopts a one-to-many matching scheme that allows multiple queries to learn from the same ground-truth event, improving coverage of ambiguous temporal regions. The second stage introduces a diversity loss ($\mathcal{L}_{dis}$) that enforces dissimilarity among query features, thereby reducing redundancy and promoting diverse representations critical for unknown event detection.}
    \label{fig:ow-sedt}
\end{figure}

We propose 1D Deformable DETR, a novel end-to-end architecture that extends Deformable DETR to the audio domain for temporal event detection. While the original Deformable DETR is designed for 2D spatial object detection using multi-scale deformable attention, our model adapts the transformer to a temporal 1D setting, using a simplified one-level Temporal Deformable Attention module. This transformation is essential for audio sequences, which inherently lie on a 1D temporal axis, unlike 2D spatial inputs in vision. An overview of the proposed WOOT model architecture is illustrated in Figure \ref{fig:ow-sedt}.

\subsubsection{Backbone}
In our framework, ResNet-50 \cite{resnet} is employed as the feature extraction backbone, motivated by its proven effectiveness in audio classification tasks and its strong capability to capture discriminative time-frequency representations \cite{hershey2017cnn, chen2019endtoendaudioclassificationbased}. Given an input audio segment, its Mel-spectrogram representation is first computed as  $X \in \mathbb{R}^{1 \times T_0 \times F_0}$, where $T_0$ and $F_0$ denote the initial temporal and frequency resolutions, respectively.  
Passing $X$ through the backbone produces a high-level feature tensor $f \in \mathbb{R}^{\mathcal{C} \times T \times F}$, with $\mathcal{C}$ representing the number of output channels and $(T, F)$ corresponding to the transformed temporal and frequency dimensions. To match the channel dimension with the transformer attention embedding size $d$, a $1 \times 1$ convolution is applied, resulting in  $z_0 \in \mathbb{R}^{d \times T \times F}$. In the conventional Deformable DETR, designed for image inputs, the CNN feature map has the shape $(B, d, H, W)$, where $B$ is the batch size, $d$ is the embedding dimension, and $(H, W)$ represents the spatial dimensions. For audio-based adaptation, this representation is transformed into a one-dimensional temporal sequence by reshaping the CNN features into $(B, T, d \times F)$. This yields the sequence  $X_S \in \mathbb{R}^{T \times D}, \quad \text{where} \quad D = d \times F$, which serves as the input to the subsequent transformer encoder.

\subsubsection{Positional encoding}

Since localization in the proposed setting is defined along a single axis, a one-dimensional sinusoidal positional encoding is employed and replicated across the frequency dimension. For a given time index $t$ and frequency bin $f$, the positional encoding is defined as

\begin{align}
P_{t,f,2i}   &= \sin\left({t}/{10000^{2i/d}}\right), \\
P_{t,f,2i+1} &= \cos\left({t}/{10000^{2i/d}}\right),
\end{align}
where $i \in \{0, 1, ..., [d/2] -1 \}$ indexes the sinusoidal pairs along the channel dimension and $d$ represents the transformer embedding size. Notably, the encoding depends solely on the temporal index $t$, and the same temporal code is broadcast uniformly across all frequency bins $f$. The resulting positional encoding tensor has the shape $P \in \mathbb{R}^{d \times T \times F}$, which is subsequently reshaped to $P \in \mathbb{R}^{T \times D}, \quad \text{where} \quad D = d \times F$, so as to match the dimensionality of $X_S$ before being added to the input feature sequence.

\subsubsection{Encoder}
Let $X_E^0 = X_S + P \in \mathbb{R}^{T \times D}$ denote the feature sequence enhanced with position information. The encoder comprises $L_E$ identical layers indexed by $\ell \in \{1, ..., L_E\}$. Each layer contains: (i) a 1D Deformable Self-Attention (1D-DSA) sub-layer and (ii) a position-wise feed-forward network (FFN). We use residual connections and LayerNorm (LN) after each sub-layer. 

\begin{equation}
\tilde{X} = \mathrm{LN}\!\big(X_E^{(\ell-1)} + \text{1D-DSA}(X_E^{(\ell-1)})\big)
\end{equation}
\begin{equation}
X_E^{(\ell)} = \mathrm{LN}\!\big(\tilde{X} + \mathrm{FFN}(\tilde{X})\big)
\end{equation}

1D Deformable Self Attention replaces dense attention with sparse sampling around a reference time, improving locality sensitivity and reducing computation compared to full attention. 1D-DSA attends only to a small set of key temporal positions around a reference point for each query, enabling the model to concentrate on the most relevant context. This is particularly suitable for audio, where important cues (e.g., transient onsets or short acoustic events) are localized and may be masked by long stretches of irrelevant background if treated with uniform attention. This 1D formulation reduces complexity, improves convergence, and makes the attention mechanism more interpretable in the audio domain.

For the reference coordinate $r_q$, a center-based normalization is used:
\begin{equation}
    r_q = (t_q + 0.5) / T
\end{equation}
where $t_q$ is the discrete temporal index of the query position $q$ in the input sequence and $T$ is the sequence length. The $+0.5$ term shifts from the left edge of the time bin to its center, ensuring symmetric sampling in both directions and avoiding boundary bias. 

The normalized sampling location for head m, query q, point j is: 
\begin{equation}
    p_{mqj} = r_q + \frac{\Delta t_{mqj}}{T}
\end{equation}
with $\Delta t_{mqj}$ are learnable offsets relative to $t_q$.

Let $z_q \in \mathbb{R}^D$ be the query feature and $r_q$ be its normalized temporal reference coordinate. For a feature sequence $\mathbf{X} \in \mathbb{R}^{T \times D}$, the output of the $m$-th attention head is computed as: 
\begin{equation}
h_m = \sum_{j=1}^J a_{mqj} W^V_mX\left(p_{mqj} \cdot T \right)
\end{equation}
with $W^V_m \in \mathbb{R}^{D \times (D / M)}$ is learnable weights, $J$ is the number of sampled points, $a_{mqj}$ are normalized attention weights. The feature $X(\cdot)$ is interpolated when the sampling location is fractional. The attention weights and offsets are predicted from $z_q$ via linear layers, with the weights normalized by \textit{softmax} so that $\sum_{j=1}^K a_{mqj} = 1$. The final head output is obtained by concatenating $h_m$ from all $M$ heads and projecting with a learnable matrix $W^O \in \mathbb{R}^{D \times D}$. 
\begin{equation}
    \text{1D-DSA}(\boldsymbol{z}_q, r_q, X) = W^{O} \text{Concat}(h_1, h_2, ..., h_m),
\end{equation}
By learning to focus on a sparse set of informative temporal positions, 1D-DSA achieves both computational efficiency and better event localization, resulting in strong performance for detecting audio events.

\subsubsection{Decoder}
The decoder receives the encoder features $X_E$ together with $N_q$ learned query embeddings as input. These $N_q$ query embeddings are learnable event slots, each representing a potential sound event instance. In the first decoding layer, they are not tied to any specific time or class, but serve as trainable probes that attend to the encoder features and progressively specialize through deformable cross-attention. The decoder is composed of $L_D$ identical blocks, each comprising a Multi-Head Self-Attention (MHSA) layer, a 1D Deformable Cross-Attention (1D-DCA) layer, and a feed-forward network. Its primary role is to generate refined event representations. As in the encoder, residual connections are applied after each sub-layer, followed by layer normalization. The principal distinction from the encoder lies in the incorporation of the 1D-DCA module. Unlike the 1D-DSA, which captures dependencies within the same input sequence, the 1D-DCA mechanism attends to the encoder features $X_E$ using the query embeddings. This layer operates on three inputs: the query embeddings from the preceding decoder layer, the coordinates of reference points within the encoder feature sequence, and the encoder features $X_E$ themselves. Through this design, each query is guided by its learnable reference points to focus selectively on a sparse set of temporal positions in the encoder output. As decoding progresses, these queries iteratively refine their attention, enabling the formation of precise and discriminative event representations for the final prediction stage.

\subsubsection{Prediction Heads}
The prediction stage converts the event representations produced by the decoder into temporal boundaries and class labels using feed-forward networks specialized for each output type. For temporal localization, each predicted event is represented by two normalized values: the temporal center and the duration. A multi-layer perceptron (MLP) is employed to obtain the timestamp. For event classification, two outputs are produced. First, a linear projection followed by a softmax activation estimates the categorical distribution over all classes. Second, the eventness head, which is modeled using a multivariate Gaussian distribution (as described in Section~\ref{cls_pred}), estimates the probability that a given query corresponds to a genuine event rather than background noise. The final classification score for each query is computed as the product of these two outputs. 

\subsection{WOOT Framework}
\subsubsection{Base Architecture} \label{cls_pred}
Our proposed framework is built upon PROB\cite{prob} as it achieves the best performance compared to other baselines. The core idea is to decouple the prediction into two parts: a classification head $f_{\text{cls}}^t(q)$ that predicts the class assuming an event is present, and an eventness head $f_{\text{event}}^t(q)$ that predicts the probability of the query containing an event. The final prediction is their product:
\begin{equation}
p(l|q) = f_{\text{cls}}^t(q) \cdot f_{\text{event}}^t(q).
\label{eq:prob}
\end{equation}

To achieve this, the distribution of all queries is modeled with a single, class-agnostic multivariate Gaussian distribution, $\mathcal{N}(\mu, \Sigma)$. The eventness score for a query $q$ is then calculated from its Mahalanobis distance $d_M(q)$ to the center of this distribution:
\begin{equation}
f_{\text{event}}^t(q)
= \exp\!\left( - (q - \mu)^\top \Sigma^{-1} (q - \mu) \right)
= \exp\!\left(-d_M(q)^2\right).
\end{equation}

The model is trained by estimating the Gaussian parameters ($\mu, \Sigma$) from queries in each batch. The eventness loss penalizes the distance of queries matched to ground-truth objects, encouraging them to be near the center of the learned object distribution. The eventness loss is defined as:
\begin{equation}
\mathcal{L}_e = \sum_{i \in \mathcal{Z}} d_M(q_i)^2.
\end{equation}
where $\mathcal{Z}$ is the set of indices for matched queries.

The PROB loss function is a weighted sum of three components: the classification loss $\mathcal{L}_{cls}$, the temporal localization loss $\mathcal{L}_{loc}$, and the eventness loss $\mathcal{L}_e$:
\begin{equation}
\mathcal{L}_{prob} = \mathcal{L}_{cls} + \mathcal{L}_{loc} + \lambda_e \mathcal{L}_e.
\end{equation}
where $\lambda_e$ is a hyperparameter balancing the eventness contribution. The classification loss $\mathcal{L}_{cls}$ is computed as the standard cross-entropy loss between the ground-truth labels and the predictions. Meanwhile, the temporal localization loss $\mathcal{L}_{loc}$ is defined as a weighted combination of an L1 regression loss and an IoU-based loss:
\begin{equation}
\mathcal{L}_{loc}= \lambda_{L1}\mathcal{L}_{L1}+\lambda_{IOU}\mathcal{L}_{IOU}.
\end{equation}

Finally, to mitigate forgetting when learning new tasks, a replay buffer is employed to retain knowledge of previously seen events. A balanced exemplar set is maintained, and the model undergoes fine-tuning on this set following each incremental training phase. At all times, at least $N_{\text{ex}}$ instances per class are preserved in the buffer to maintain performance on earlier tasks.

\subsubsection{Feature Disentangle}
A key challenge in open-world sound event detection is that features useful for identifying whether a region contains any event are often entangled with class-specific cues required for classification. This coupling causes two main issues. First, it limits the model's ability to generalize to unknown events, since the representation is overly tied to known class identities. Second, it harms incremental learning, since updating class-specific features can overwrite those used for detecting past or unseen events.

To address this, we propose to disentangle the feature of each object query into two components: a \textit{class-agnostic} feature and a \textit{class-specific} feature. The class-agnostic feature is used for computing the eventness loss, while the class-specific feature is used for classification. The original query feature is retained for temporal localization.

To achieve this disentanglement, we introduce a disentanglement block composed of multiple linear layers. Given a query embedding $q$, we compute the class-agnostic feature $q_\text{agn}$ as:
\begin{equation}
q_\text{agn} = f_{\text{dis}}(q),
\end{equation}
where $f_{\text{dis}}(\cdot)$ denotes the disentanglement block. The class-specific feature $q_\text{spec}$ is then obtained by subtracting the agnostic component from the original:
\begin{equation}
q_\text{spec} = q - q_\text{agn}.
\end{equation}

To further encourage the independence between class-agnostic feature and class-specific feature, we add a disentangle loss, which is defined as:

\begin{equation}
\mathcal{L}_{dis} = \frac{1}{N} \sum_{i=1}^{N} \left| \frac{q_{\text{agn}}^{(i)} \cdot q_{\text{spec}}^{(i)}}{\|q_{\text{agn}}^{(i)}\| \, \|q_{\text{spec}}^{(i)}\|} \right|.
\end{equation}

where \( N \) is the total number of object queries.

This design allows the model to independently model event presence and class identity. It not only improves detection of unknown events by focusing on generalizable features but also reduces forgetting in incremental learning by reducing interference from new class updates and helping retain prior knowledge.

\subsubsection{Two-Stage Training Strategy} Unlike in object detection, where detecting only part of an object is typically not meaningful, in sound event detection, even partial segments of an event often still belong to the same class. However, conventional one-to-one matching strategies assign only one query per ground-truth event, potentially discarding other valid queries that partially or fully cover the same event. For example, a query whose predicted segment lies entirely within a ground-truth interval might be ignored due to a higher localization cost.

To tackle this issue, we employ a one-to-many matching strategy. Specifically, after performing standard bipartite matching, the matched queries are considered \textit{fully-matched}. We then additionally consider unmatched queries as \textit{semi-matched} if they satisfy two conditions: (1) the predicted class confidence is greater than a threshold $\alpha$, and (2) the ratio of its intersection with a same-class ground-truth segment to its own predicted segment length is greater than a threshold $\beta$. The \textit{semi-matched} queries are then used for training in the same way as \textit{fully-matched} queries, except that their localization loss is set to $0$.

However, this one-to-many matching strategy introduces a new issue: multiple queries may be optimized to represent the same ground-truth event, leaving fewer queries available to represent unknown events. Moreover, unmatched queries intended to capture unknown events may redundantly focus on different segments of the same unknown event. To mitigate these problems, we propose a diversity loss that encourages the unmatched queries to produce diverse representations. The diversity loss is defined as:
\begin{equation}
\mathcal{L}_{div} = \frac{1}{|\mathcal{Q}_{um}|(|\mathcal{Q}_{um}| - 1)} \sum_{\substack{i, j \in \mathcal{Q}_{um} \\ i \neq j}} \left( \frac{q_i \cdot q_j}{\|q_i\| \, \|q_j\|} \right).
\end{equation}
where $\mathcal{Q}_{um}$ is the set of unmatched queries.

To allow the model to fully exploit the \textit{semi-matched} queries, we do not apply the diversity loss during the initial training phase (referred to as \textbf{stage 1}). After a certain number of epochs, we enter \textbf{stage 2}, where the diversity loss is applied to promote diversity among unmatched queries.

To summarize, our loss function is:
\begin{equation}
\mathcal{L}_{total} = \mathcal{L}_{cls} + \mathcal{L}_{loc} + \lambda_e \mathcal{L}_e + \lambda_{dis}\mathcal{L}_{dis} + \lambda_{div} \mathcal{L}_{div}.
\end{equation}
where in stage $1$, $\lambda_{div}$ equals to $0$.

\section{Experiments and Results}
\subsection{Open-World Evaluation Protocol}
\subsubsection{Data split}

The full set of classes is partitioned into distinct, non-overlapping tasks $\{T_1, \cdots, T_t, \cdots\}$, where each group $T_\tau$ becomes available only at time step $t = \tau$. During training on task $T_t$, classes introduced in all previous and current tasks $\{T_\tau \mid \tau \leq t\}$ are treated as known, while those from future tasks $\{T_\tau \mid \tau > t\}$ remain unknown. Specifically, we split classes of URBAN-SED~\cite{urban-sed} and DESED~\cite{desed} dataset into 3 tasks, as shown in Table~\ref{tab:data_split} and Table~\ref{tab:data_split_desed} . We use the training set of each dataset for training, while evaluation is performed on the URBAN-SED test set and the DESED public evaluation set.

\begin{table} [t]
\centering
\caption{Task configuration and data split statistics for the URBAN-SED dataset.}
\resizebox{\textwidth}{!}{%
\begin{tabular}{@{}l@{\hskip 2pt}|@{\hskip 5pt}c@{\hskip 5pt}c@{\hskip 5pt}c@{}} 
\toprule
& Task 1 & Task 2 & Task 3  \\ \midrule
Class split & \begin{tabular}[c]{@{}c@{}}air conditioner, engine\\idling, jackhammer\end{tabular} 
            & \begin{tabular}[c]{@{}c@{}}street music, dog\\bark, children playing\end{tabular} 
            & \begin{tabular}[c]{@{}c@{}}car horn, drilling,\\gun shot, siren\end{tabular} \\ \midrule
\# training audios & 4493 & 4469 & 5006 \\
\# test audios     & 1472 & 1483 & 1659 \\
\# train instances & 7865 & 7872 & 10705 \\
\# test instances  & 2562 & 2620 & 3620 \\

\bottomrule
\end{tabular}%
}
\vspace{0.5mm}
\label{tab:data_split}
\end{table}

\begin{table} [t]
\centering
\caption{Task configuration and data split statistics for the DESED dataset.}
\resizebox{\textwidth}{!}{%
\begin{tabular}{@{}l@{\hskip 2pt}|@{\hskip 5pt}c@{\hskip 5pt}c@{\hskip 5pt}c@{}} 
\toprule
& Task 1 & Task 2 & Task 3  \\ \midrule
Class split & \begin{tabular}[c]{@{}c@{}}alarm bell ringing, cat,\\running water, vacuum cleaner\end{tabular} 
            & \begin{tabular}[c]{@{}c@{}}dishes, dog,\\ frying\end{tabular} 
            & \begin{tabular}[c]{@{}c@{}}speech, blender,\\electric shaver toothbrush\end{tabular} \\ \midrule
\# training audios & 4746 & 3956 & 9339 \\
\# test audios     & 332 & 252 & 361 \\
\# train instances & 6117 & 7976 & 18003 \\
\# test instances  & 641 & 1019 & 1105 \\

\bottomrule
\end{tabular}%
}
\vspace{0.5mm}
\label{tab:data_split_desed}
\end{table}

\subsubsection{Evaluation Metric}
There are two popular ways for evaluating sound event detection systems: event-based and segment-based. Among these, the event-based metric is more suitable for open-world SED, as it directly evaluates the ability to localize and classify individual sound events in time, including unseen categories. In contrast, the segment-based metric only measures whether each class is active within a segment. In the open-world setting, the ``unknown'' label may correspond to many different actual classes. If multiple unknown events with different ground-truth labels occur in the same segment, correctly detecting only one of them is sufficient for the segment-based metric to count the prediction as correct, thereby masking missed detections of other unknown events. Therefore, we adopt the event-based F1-score for known sound event detection.

For unknown sound event detection, we use event-based macro recall as the main metric, as the dataset does not include annotations for every potential unknown event. Using recall under such conditions is consistent with prior studies~\cite{prob,cat,joseph2021towards}.
\subsection{Implementation Details}
All experiments were conducted using two NVIDIA RTX 6000 GPUs, each equipped with 24 GB of VRAM. In each task, the WOOT framework is first trained for $200$ epochs, followed by an additional $200$ epochs of fine-tuning during the incremental learning step. The final $100$ epochs of each task are designated as stage~$2$ training. Models are optimized using the AdamW optimizer with a batch size of $128$, an initial learning rate of $10^{-4}$, and a weight decay of $10^{-4}$. The learning rate is decreased by a factor of $10$ during the fine-tuning phase of the incremental step. The number of queries is set to $18$ in all experiments. Our model contains approximately 37.4M trainable parameters. For the loss coefficients, we set $\lambda_{L1}=5, \lambda_{IOU}=2, \lambda_{e} = 8\times 10^{-4},\ \lambda_{dis} = 10^{-3},\ \lambda_{div} = 10^{-2}$.  For the exemplar replay, we choose $N_{ex} = 200$.

\subsection{Open-World Sound Event Detection Results}
\begin{table*}[b]
\centering
\caption{Performance comparison of WOOT and state-of-the-art OWOD-derived methods on URBAN-SED. ``-'' indicates that the metric is not applicable to the method. In Task 3, U-Recall is not computed, as it contains only known classes.}
\label{tab:sota_owod}
\setlength{\tabcolsep}{3pt}
\adjustbox{width=\textwidth}{
\begin{tabular}{@{}l|cc|cccc|ccc@{}}
\toprule
\textbf{Task IDs} ($\rightarrow$) & \multicolumn{2}{c|}{\textbf{Task 1}} & \multicolumn{4}{c|}{\textbf{Task 2}} & \multicolumn{3}{c}{\textbf{Task 3}} \\ \midrule
 & \cellcolor[HTML]{FFFFED}U-Recall ($\uparrow$) & \cellcolor[HTML]{EDF6FF}F1 ($\uparrow$) 
 & \cellcolor[HTML]{FFFFED}U-Recall ($\uparrow$) & \multicolumn{3}{c|}{\cellcolor[HTML]{EDF6FF}F1 ($\uparrow$)} 
 & \multicolumn{3}{c}{\cellcolor[HTML]{EDF6FF}F1 ($\uparrow$)} \\
 &  & Cur known &  & Prev known & Cur known & Both & Prev known & Cur known & Both \\ \midrule

\begin{tabular}[c]{@{}l@{}}1D DETR~\cite{sedt}\\\hspace{1em}+ Finetuning\end{tabular} 
& - & $34.7^{\pm 0.5}$ & - & $11.4^{\pm 0.4}$ & $20.7^{\pm 0.5}$ & $15.8^{\pm 0.6}$ & $7.9^{\pm 0.3}$ & $28.9^{\pm 0.4}$ & $16.3^{\pm 0.3}$ \\

\begin{tabular}[c]{@{}l@{}}Ours: 1D DDETR \\\hspace{1em}+ Finetuning\end{tabular} 
& - & $47.1^{\pm 0.5}$ & - &  $21.2^{\pm 0.8}$ & $28.4^{\pm 0.4}$ &  $24.8^{\pm 0.3}$ & $15.5^{\pm 0.3}$ & $36.5^{\pm 0.8}$ & $23.9^{\pm 0.4}$ \\

\midrule

OW-DETR~\cite{ow-detr} &\cellcolor[HTML]{FFFFED}$18.8^{\pm 0.5}$ & $43.1^{\pm 0.1}$ &\cellcolor[HTML]{FFFFED}$25.8^{\pm 0.7}$ & $16.7^{\pm 0.6}$ & $25.0^{\pm 0.5}$ & $20.9^{\pm 0.6}$ & $12.6^{\pm 1.1}$ & $33.8^{\pm 0.7}$ & $21.2^{\pm 0.9}$ \\
SS OW-DETR~\cite{ss-owdetr} &\cellcolor[HTML]{FFFFED}$15.5^{\pm 0.6}$  & $42.3^{\pm 0.8}$ &\cellcolor[HTML]{FFFFED}$20.8^{\pm 0.8}$  & $15.6^{\pm 0.6}$ & $24.2^{\pm 0.7}$ & $19.3^{\pm 0.6}$ & $12.3^{\pm 0.5}$ & $33.7^{\pm 0.5}$ & $20.7^{\pm 0.4}$ \\
PROB~\cite{prob} (Baseline)
& \cellcolor[HTML]{FFFFED}$21.4^{\pm 0.4}$
& $46.1^{\pm 0.5}$
& \cellcolor[HTML]{FFFFED}$27.7^{\pm 0.8}$
& $18.2^{\pm 0.9}$
& $25.3^{\pm 0.6}$
& $21.6^{\pm 0.6}$
& $15.1^{\pm 0.7}$
& $35.3^{\pm 0.5}$
& $23.2^{\pm 0.5}$ \\
CAT~\cite{cat} &\cellcolor[HTML]{FFFFED}$19.5^{\pm 0.8}$  & $45.1^{\pm 0.5}$ &\cellcolor[HTML]{FFFFED}$29.3^{\pm 0.9}$  & $18.0^{\pm 1.2}$ & $22.8^{\pm 0.3}$ & $21.5^{\pm 0.7}$ & $14.8^{\pm 0.8}$ & $36.2^{\pm 0.7}$ & $23.3^{\pm 0.3}$ \\

\midrule

\textbf{Ours: WOOT} 
& \cellcolor[HTML]{FFFFED}$\mathbf{28.6^{\pm 0.5}}$ & $\mathbf{48.4^{\pm 0.1}}$ 
& \cellcolor[HTML]{FFFFED}$\mathbf{33.4^{\pm 0.3}}$ & $\mathbf{23.5^{\pm 0.4}}$ & $\mathbf{25.9^{\pm 0.4}}$ & $\mathbf{24.6^{\pm 0.5}}$ & $\mathbf{17.1^{\pm 0.8}}$ & $\mathbf{34.5^{\pm 0.9}}$ & $\mathbf{24.1^{\pm 0.2}}$ \\

\bottomrule
\end{tabular}%
}
\vspace{-0.2cm}
\end{table*}

In this section, a series of detailed experiments is conducted to evaluate the performance of the proposed WOOT framework. Our method is compared against leading approaches in the OWOD domain. Table \ref{tab:sota_owod} presents the comparative results based on the Unknown Class Recall (U-Recall) and the F1 score for Known Classes. To ensure robustness, all experiments are repeated with multiple random seeds, and we report the mean performance along with the standard deviation.

First, the performance of the proposed 1D Deformable DETR architecture is compared with that of the standard 1D DETR model introduced in \cite{sedt} for the sound event detection task across the three datasets described in the preceding section (Task 1, Task 2, and Task 3). For a fair comparison, both models are fine-tuned on Tasks 2 and 3 to mitigate catastrophic forgetting. Since 1D DETR and 1D DDETR are designed solely for classifying events into known categories, they cannot accommodate unknown class detection, and therefore, U-Recall is not applicable to them. Based on the results in Table \ref{tab:sota_owod}, the 1D DDETR consistently outperforms 1D DETR across all evaluation metrics. In Task 1, the improvement reaches nearly $13\%$, while for the remaining tasks, the F1 score of 1D DDETR is generally higher by approximately $8$ - $9\%$. These results demonstrate the superior capability of the proposed architecture in modeling temporal dependencies and capturing discriminative features, thereby yielding more accurate event localization and classification compared to the conventional 1D DETR.

With the strong performance of the 1D DDETR architecture, all subsequent experiments incorporating Open World Learning techniques were conducted using this architecture to ensure fairness in comparison. State-of-the-art methods in the OWOD domain were evaluated. These methods can be broadly grouped into two categories: \textbf{(i)} Pseudo-labeling based approaches, including OW-DETR \cite{ow-detr}, SS OW-DETR \cite{ss-owdetr}, and CAT \cite{cat}, which rely on iterative refinement of pseudo-labels to handle unknown instances; and \textbf{(ii)} Probabilistic frameworks for objectness estimation, represented by PROB \cite{prob}, which models objectness as a probabilistic variable to improve classification confidence for both known and unknown classes. Owing to the differences between the OWOD and OW-SED tasks, the core OWL techniques proposed in each work were adapted and integrated into the 1D DDETR architecture. Specifically, for OW-DETR, the attention-driven pseudo-labeling, novel classification layer, and objectness scoring modules were adopted, with the objectness scoring mechanism modified to operate over the temporal dimension instead of the 2D spatial domain. For SS OW-DETR, the Object Query Guided Pseudo-Labeling strategy was also adapted to the one-dimensional setting. In the case of CAT, both the Cascade Decoupled Decoding and the Self-Adaptive Pseudo-Labeling Mechanism were fully incorporated. Similarly, for PROB, the Probabilistic Objectness module was applied in its entirety to classify events. The detailed results of these experiments are presented in the lower part of Table~\ref{tab:sota_owod}. 


Within the pseudo-labeling group, CAT achieves the best overall performance. It attains $19.5\pm0.8$ U-Recall in Task 1 and $29.3\pm0.9$ in Task 2, surpassing OW-DETR ($18.8\pm0.5,\,25.8\pm0.7$) and SS OW-DETR ($15.5\pm0.6,\,20.8\pm0.8$). Its known-class F1 scores remain competitive, reaching $21.5\pm0.7$ (Both) in Task 2 and $23.3\pm0.3$ (Both) in Task 3. PROB, despite belonging to a different methodological category, delivers comparable results and remains a strong baseline: it achieves the highest U-Recall among prior works in Task 1 ($21.4\pm0.4$), and shows consistently strong F1 across both previous and current known classes (e.g., $18.2\pm0.9$ and \ $25.3\pm0.6$ in Task 2; $15.1\pm0.7$ and\ $35.3\pm0.5$ in Task 3). This suggests that probabilistic modeling of objectness is a compelling alternative to pseudo-labeling for open-world scenarios.

While CAT and PROB establish strong baselines within their respective methodological categories, the proposed WOOT pushes the performance boundary even further. Our framework surpasses all baselines across most metrics, with especially pronounced improvements in unknown event detection. WOOT achieves $28.6\pm0.5$ U-Recall in Task 1 and $33.4\pm0.3$ in Task 2, corresponding to improvements of +$9.1$ and +$4.1$ over CAT, and +$7.2$ and +$5.7$ over PROB, the previous best performers for that metric. These correspond to relative improvements of approximately $33.6$ (Task 1) and $14.0$ (Task 2) compared to the strongest baseline in each case. Notably, these gains are achieved while retaining high accuracy on known classes. Moreover, to explicitly report catastrophic forgetting, we quantify forgetting between two consecutive tasks $t-1 \to t$ as the drop in performance on the classes learned in the previous task after adapting to the new task:
\begin{equation}
    \mathcal{F}_{t-1 \to t} = F1_{cur}(t-1) - F1_{prev}(t)
\end{equation}
where lower values indicate better retention. Under this definition, WOOT exhibits reduced forgetting from Task 1 to Task 2, achieving $\mathcal{F}_{1 \to 2} = 48.4 - 23.5=24.9$, which is the lowest among all compared open-world frameworks (OW-DETR: 26.4, SS OW-DETR: 26.7, PROB: 27.9, CAT: 27.1), while simultaneously attaining the highest retained performance on old classes after Task 2 ($23.5\pm0.4$). From Task 2 to Task 3, WOOT incurs a $8.8$ drop ($25.9 - 17.1$) yet still preserves the strongest performance on old classes in Task~3 ($17.1\pm0.8$), reflecting robust resistance to catastrophic forgetting under continued class expansion.

\begin{table*}[t]
\centering
\caption{Performance comparison of WOOT and state-of-the-art OWOD-derived methods on DESED.}
\label{tab:sota_owod_desed}
\setlength{\tabcolsep}{3pt}
\adjustbox{width=\textwidth}{
\begin{tabular}{@{}l|cc|cccc|ccc@{}}
\toprule
\textbf{Task IDs} ($\rightarrow$) & \multicolumn{2}{c|}{\textbf{Task 1}} & \multicolumn{4}{c|}{\textbf{Task 2}} & \multicolumn{3}{c}{\textbf{Task 3}} \\ \midrule
 & \cellcolor[HTML]{FFFFED}U-Recall ($\uparrow$) & \cellcolor[HTML]{EDF6FF}F1 ($\uparrow$) 
 & \cellcolor[HTML]{FFFFED}U-Recall ($\uparrow$) & \multicolumn{3}{c|}{\cellcolor[HTML]{EDF6FF}F1 ($\uparrow$)} 
 & \multicolumn{3}{c}{\cellcolor[HTML]{EDF6FF}F1 ($\uparrow$)} \\
 &  & Cur known &  & Prev known & Cur known & Both & Prev known & Cur known & Both \\ \midrule

\begin{tabular}[c]{@{}l@{}}1D DETR~\cite{sedt}\\\hspace{1em}+ Finetuning\end{tabular} 
& - & $21.4^{\pm 0.5}$ & - & $14.3^{\pm 0.7}$ & $18.6 ^{\pm 0.9}$ & $16.2^{\pm 0.8}$ & $10.4^{\pm 0.7}$ & $14.2^{\pm 0.6}$ & $11.4^{\pm 0.4}$ \\

\begin{tabular}[c]{@{}l@{}}Ours: 1D DDETR \\\hspace{1em}+ Finetuning\end{tabular} 
& - & $33.8^{\pm 0.8}$ & - & $30.6^{\pm 0.8}$ & $28.6^{\pm 1.7}$ &$29.7^{\pm 1.1}$ & $16.0^{\pm 1.2}$ & $22.1^{\pm 1.7}$ & $17.9^{\pm 0.5}$ \\

\midrule

OW-DETR~\cite{ow-detr} &\cellcolor[HTML]{FFFFED}$10.8^{\pm 1.4}$ & $29.0^{\pm 0.5}$ &\cellcolor[HTML]{FFFFED}$11.2^{\pm 0.7}$ & $24.7^{\pm 1.2}$ & $26.0^{\pm 0.4}$ & $25.2^{\pm 1.2}$ & $12.6^{\pm 0.9}$ & $15.1^{\pm 0.4}$ & $14.0^{\pm 0.5}$ \\
SS OW-DETR~\cite{ss-owdetr} &\cellcolor[HTML]{FFFFED}$7.7^{\pm 0.9}$  & $29.3^{\pm 0.5}$ &\cellcolor[HTML]{FFFFED}$10.0^{\pm 0.7}$  & $24.8^{\pm 0.4}$ & $24.5^{\pm 1.1}$ & $24.7^{\pm 0.9}$ & $11.5^{\pm 0.8}$ & $14.1^{\pm 0.6}$ & $12.4^{\pm 0.6}$ \\
PROB~\cite{prob} (Baseline)
& \cellcolor[HTML]{FFFFED}$15.5^{\pm 0.3}$ & $31.0^{\pm 0.8}$ & \cellcolor[HTML]{FFFFED}$12.4^{\pm 0.4}$ & $28.1^{\pm 0.5}$ & $27.7^{\pm 0.4}$& $27.9^{\pm 0.2}$ & $16.3^{\pm 0.7}$ & $23.6^{\pm 0.7}$ & $18.5^{\pm 0.7}$\\
CAT~\cite{cat} &\cellcolor[HTML]{FFFFED}$13.0^{\pm 0.6}$  & $30.8^{\pm 0.6}$ &\cellcolor[HTML]{FFFFED}$13.2^{\pm 0.6}$  & $23.7^{\pm 1.3}$ & $26.2^{\pm 0.4}$ & $24.5^{\pm 1.8}$ & $15.5^{\pm 0.7}$ & $18.8^{\pm 1.2}$ & $16.6^{\pm 1.0}$ \\

\midrule

\textbf{Ours: WOOT} 
& \cellcolor[HTML]{FFFFED}$\mathbf{18.3^{\pm 0.2}}$ & $\mathbf{32.5^{\pm 0.7}}$ 
& \cellcolor[HTML]{FFFFED}$\mathbf{14.0^{\pm 0.3}}$ & $\mathbf{30.4^{\pm 0.6}}$ & $\mathbf{28.1^{\pm 0.5}}$ & $\mathbf{29.4^{\pm 0.2}}$ &  $\mathbf{17.0^{\pm 0.5}}$ &  $\mathbf{25.7^{\pm 0.1}}$ &  $\mathbf{19.6^{\pm 0.4}}$ \\

\bottomrule
\end{tabular}%
}
\vspace{-0.2cm}
\end{table*}

A similar performance trend is observed on the DESED dataset, as shown in Table \ref{tab:sota_owod_desed}. Our proposed WOOT framework achieves superior performance across all metrics compared to established baselines. Specifically, WOOT improves the U-Recall by approximately 18\% in Task 1 and 13\% in Task 2 relative to the strongest previous method (PROB). Furthermore, our approach demonstrates a strong ability to retain knowledge of previously known classes, achieving the highest F1 scores of $30.4\pm0.6$ in Task 2 and $17.0\pm0.5$ in Task 3. These results confirm that the effectiveness of our approach generalizes well across different acoustic environments and is not limited to a single dataset.

\begin{table*}[h]
\centering
\caption{Effect of gradually incorporating our proposed components into the baseline model. “FD” denotes feature disentangle and “TSTS” indicates two-stage training strategy. Results are averaged over multiple random seeds (mean).}
\label{tab:ablation_component}
\setlength{\tabcolsep}{3pt}
\adjustbox{width=\textwidth}{
\begin{tabular}{@{}l|cc|cccc|ccc@{}}
\toprule
\textbf{Task IDs} ($\rightarrow$) & \multicolumn{2}{c|}{\textbf{Task 1}} & \multicolumn{4}{c|}{\textbf{Task 2}} & \multicolumn{3}{c}{\textbf{Task 3}} \\ \midrule
 & \cellcolor[HTML]{FFFFED}U-Recall ($\uparrow$) & \cellcolor[HTML]{EDF6FF}F1 ($\uparrow$) 
 & \cellcolor[HTML]{FFFFED}U-Recall ($\uparrow$) & \multicolumn{3}{c|}{\cellcolor[HTML]{EDF6FF}F1 ($\uparrow$)} 
 & \multicolumn{3}{c}{\cellcolor[HTML]{EDF6FF}F1 ($\uparrow$)} \\
 &  & Cur known &  & Prev known & Cur known & Both & Prev known & Cur known & Both \\ \midrule

Baseline
& \cellcolor[HTML]{FFFFED}21.4 & 46.1 & \cellcolor[HTML]{FFFFED}27.7 & 18.2 & 25.3 & 21.6 & 15.1 & 35.3 & 23.2 \\

\midrule

\begin{tabular}[c]{@{}l@{}}Basline + TSTS\end{tabular} 
 & \cellcolor[HTML]{FFFFED}23.0 & 47.1  & \cellcolor[HTML]{FFFFED}29.4 & 19.1 & 26.4 & 22.7 & 16.2 & 34.0 & 23.5 \\

\begin{tabular}[c]{@{}l@{}}Baseline + FD \end{tabular} 
& \cellcolor[HTML]{FFFFED}25.2 & 47.3 & \cellcolor[HTML]{FFFFED}31.3 & 20.1 & 26.3 & 23.2 & 16.3 & 34.5 & 23.6 \\

Final: \textbf{WOOT} 
& \cellcolor[HTML]{FFFFED}\textbf{28.6} & \textbf{48.4} 
& \cellcolor[HTML]{FFFFED}\textbf{33.4} & \textbf{23.5} & \textbf{25.9} & \textbf{24.6} & \textbf{17.1} & \textbf{34.5} & \textbf{24.1} \\

\bottomrule
\end{tabular}%
}
\vspace{-0.2cm}
\end{table*}

\subsection{Ablation Study on WOOT framework}
\subsubsection*{Impact of core components}
An ablation analysis is performed to assess the impact of each core component independently, as well as their combined effect in our WOOT framework: the feature disentangle (FD) module and the two-stage training strategy (TSTS). Table~\ref{tab:ablation_component} presents the performance when progressively integrating these components into the baseline. 

TSTS improves both unknown-event detection and known-event recognition across all three tasks. For example, in Task 2, adding TSTS to the baseline raises unknown-event recall from $27.7$ to $29.4$, and improves current-known F1 from $25.3$ to $26.4$ and previous-known F1 from $18.2$ to $19.1$. Similar upward trends are observed in Task 1 and Task 3, confirming that TSTS’s strategy of treating more queries as positive matches provides richer learning signals, thereby enhancing both generalization to unseen classes and retention of known ones.

On the other hand, FD produces a stronger boost in unknown detection performance. In Task 1, adding FD increases unknown-event recall from $21.4$ to $25.2$, and in Task 2, it jumps from $27.7$ to $31.3$. This aligns with its purpose: FD explicitly disentangles class-agnostic “eventness” features from class-specific information, enabling the model to learn a more stable and invariant eventness representation that is particularly valuable for identifying unseen events.

When both methods are combined, their effects are cumulative. The full WOOT model achieves the highest scores in nearly all metrics. This synergy suggests that TSTS expands the amount of effective training supervision, while FD enhances the quality of the learned representations, together forming a more robust and generalizable framework.

\begin{figure*}
    \centering
    \includegraphics[width=1\linewidth]{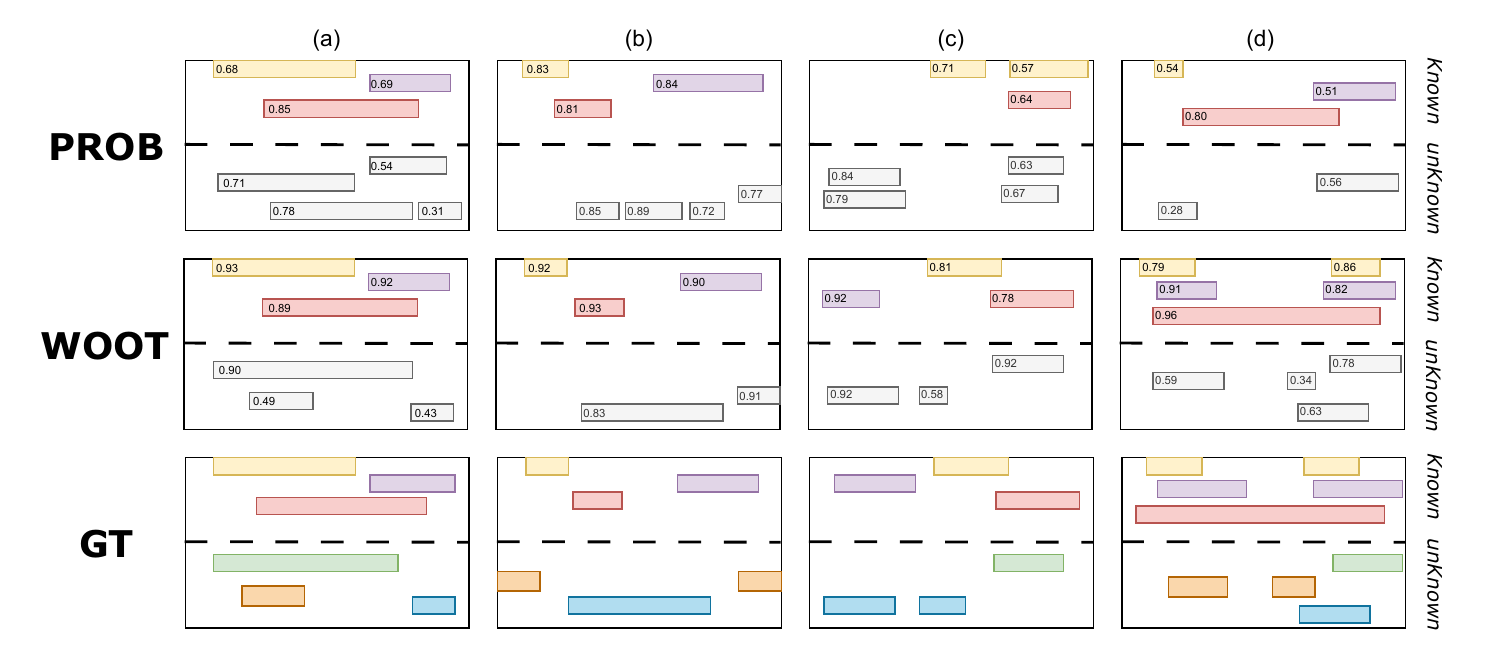}
    \caption{Visualization of the outputs from PROB and our framework compared with the ground truth (GT) in Task 1 of URBAN-SED. Each color corresponds to a different class, while grey indicates an unknown prediction. The number in each segment indicates the confidence score, computed using Equation ~\ref{eq:prob}. The dashed line in the middle separates known and unknown regions. Note that for clearer visualization, only a subset of the unknown predictions is shown.}
    \label{fig:visual}
\end{figure*}

For further understanding, we visualize some outputs of PROB, our approach, and the ground truth in Figure \ref{fig:visual}. As shown in (a), (c), and (d), the PROB outputs often produce an unknown prediction whenever a known class is detected, either as a sub-segment or as a heavily overlapping region. This happens because, in PROB’s original training, such overlapping regions were treated as unknown/background classes, and since they share high similarity with the known class predictions, they often receive high eventness scores. Together, these factors lead them to be classified as unknown. Consequently, this reduces the unknown recall by limiting the number of free queries available for detecting other unknown classes. Our approach overcomes this issue through the concept of semi-matched queries. Moreover, as shown in (b) and (c), the PROB outputs frequently contain multiple unknown predictions that should ideally be collapsed into a single larger segment or two unknown queries corresponding to the same event. In contrast, our method addresses this problem by introducing a diversity loss, which encourages different queries to focus on different unknown events. Furthermore, as shown in (c) and (d), our method not only yields more accurate known and unknown predictions but also assigns higher confidence scores to the correct predictions.

\subsubsection*{Impact of the Number of Queries}
\begin{table*}[b]
\centering
\caption{Impact of varying query numbers in the 1D-DDETR architecture of the WOOT model. Results are averaged over multiple random seeds (mean).}
\label{tab:ablation_query}
\setlength{\tabcolsep}{3pt}
\adjustbox{width=\textwidth}{
\begin{tabular}{@{}c|cc|cccc|ccc@{}}
\toprule
\multirow{3}{*}{\textbf{\#Queries}} & \multicolumn{2}{c|}{\textbf{Task 1}} & \multicolumn{4}{c|}{\textbf{Task 2}} & \multicolumn{3}{c}{\textbf{Task 3}} \\ \cmidrule{2-10}
 & \cellcolor[HTML]{FFFFED}U-Recall ($\uparrow$) & \cellcolor[HTML]{EDF6FF}F1 ($\uparrow$) 
 & \cellcolor[HTML]{FFFFED}U-Recall ($\uparrow$) & \multicolumn{3}{c|}{\cellcolor[HTML]{EDF6FF}F1 ($\uparrow$)} 
 & \multicolumn{3}{c}{\cellcolor[HTML]{EDF6FF}F1 ($\uparrow$)} \\
 ($\downarrow$) &  & Cur known &  & Prev known & Cur known & Both & Prev known & Cur known & Both \\ \midrule

$12$
 & \cellcolor[HTML]{FFFFED}$20.9$ & $47.9$ & \cellcolor[HTML]{FFFFED}$25.5$ & $22.4$ & $26.5$ & $24.4$ & $15.7$ & $33.4$ & $22.8$ \\

$18$
& \cellcolor[HTML]{FFFFED}$28.6$ & $48.4$ 
& \cellcolor[HTML]{FFFFED}$33.4$ & $23.5$ & $25.9$ & $24.6$ & $17.1$ & $34.5$ & $24.1$ \\

$24$
& \cellcolor[HTML]{FFFFED}$30.1$ & $46.4$ 
& \cellcolor[HTML]{FFFFED}$37.4$ & $23.0$ & $25.5$ & $24.3$ & $16.8$ & $34.7$ & $24.1$ \\

$30$
& \cellcolor[HTML]{FFFFED}$36.2$ & $46.3$ 
& \cellcolor[HTML]{FFFFED}$41.0$ & $22.9$ & $25.4$ & $24.0$ & $16.6$ & $34.4$ & $23.9$ \\

\bottomrule
\end{tabular}%
}
\vspace{-0.2cm}
\end{table*}

Table \ref{tab:ablation_query} presents how adjustments in the number of queries influence the proposed 1D-DDETR architecture. The goal of this experiment is to examine how query capacity influences performance across incremental tasks in the open-world SED setting. The results show that raising the number of queries generally leads to improvements in U-Recall, particularly in Task 1 (from 20.9 with 12 queries up to 36.2 with 30 queries) and Task 2 (from 25.5 to 41.0). This indicates that more queries provide greater flexibility in capturing diverse unknown events. However, the F1 score, especially for current and previous known classes, remains relatively stable or slightly decreases as queries increase, with the best balance observed at 18 queries. At this setting, U-Recall significantly improves over 12 queries (28.6 compared to 20.9 in Task 1; 33.4 compared to 25.5 in Task 2), while F1 scores remain slightly better than larger query sizes. This suggests that using 18 queries achieves a favorable trade-off, offering sufficient capacity to represent unknown events without diluting the discriminative power for known classes.

\subsection{Closed-World Sound Event Detection Results}

We also evaluate our proposed 1D-DDETR in a closed-world setting. Following previous work~\cite{sedt}, we report results for both CRNN-based and Transformer-based models (denoted as “CTrans”), as well as their variants post-processed by adaptive median filtering, referred to as CRNN-CWin and CTrans-CWin, respectively. Table~\ref{tab:urban} presents the performance in terms of Eb (event-based macro F1), Sb (segment-based macro F1), and At (audio tagging macro F1). On URBAN-SED, 1D-DDETR improves substantially on Eb over 1D-DETR ($32.71 \to 37.02$), while Sb and At are roughly comparable to 1D-DETR but remain below CRNN and CTrans baselines. This trend is consistent with the design of DETR-family detectors, which optimize a set-prediction objective with bipartite matching and therefore emphasize a compact set of unique event instances with accurate temporal boundaries. In contrast, Sb and At benefit more from dense frame or segment evidence and from modeling temporal continuity. As a result, frame-based SED architectures such as CRNN and CTrans, which are trained to produce dense time-frame predictions, often retain an advantage on Sb and At. 

\begin{table}[h]
    \centering
    \caption{Evaluation results on the URBAN-SED testing dataset}
    \begin{tabular}{cccc}
    \toprule
        Model & Eb$[\%]$ & Sb$[\%]$ & At$[\%]$\\
        \midrule
         CTrans & $31.33$ & $64.51$ & $74.67$ \\
         CTrans-CWin & $34.36$ & $64.73$ & $74.05$ \\
         \midrule
         CRNN & $35.26$ & $65.75$ & $74.64$\\
         CRNN-CWin & $36.75$ & $65.74$ & $74.19$ \\
         \midrule
         1D-DETR & $32.71$ & $60.64$ & $70.90$ \\
         1D-DDETR (Ours) & $37.02$ & $60.77$ & $71.15$\\
         \bottomrule
    \end{tabular}
    \label{tab:urban}
\end{table}

\subsection{Ablation Study on 1D-Deformable DETR}

\begin{table}[h]
\centering
\caption{Ablation study on the impact of deformable encoder and deformable decoder in 1D-Deformable DETR}.
\label{tab:ablation_defm}
\begin{tabular}{cc|ccc}
\toprule
Deformable Enc. & Deformable Dec. &  Eb$[\%]$ & Sb$[\%]$ & At$[\%]$  \\
\midrule
\ding{55} & \ding{55} & 31.12 & 59.39 & 69.16\\
\ding{51} & \ding{55} & 33.58 & 60.75 & 70.51 \\
\ding{55} & \ding{51} & 35.73 & 59.48 & 70.02 \\
\ding{51} & \ding{51} & 37.02 & 60.77 & 71.15 \\
\bottomrule
\end{tabular}
\end{table}


In this section, the impact of deformable attention is examined in comparison to standard dense attention. As shown in Table \ref{tab:ablation_defm}, the deformable attention in the are replaced encoder, decoder, and both modules by dense attention to analyze its impact. The results clearly highlight that both the deformable encoder and deformable decoder play critical roles in improving detection accuracy. This is primarily due to dense attention's inability to effectively capture local context. In audio, consecutive segments are often highly similar. Dense attention treats all positions equally, which dilutes the focus on truly informative temporal cues and increases computational burden.

\section{Conclusions}
In this study, the Open-World Sound Event Detection (OW-SED) problem was introduced, extending the open-world learning paradigm from the visual domain to audio understanding. A novel 1D Deformable architecture was proposed to address the unique temporal characteristics of sound events, allowing the model to adaptively attend to informative temporal regions while maintaining locality awareness. Building upon this backbone, an Open-World Deformable Sound Event Detection Transformer (WOOT) framework was developed that incorporates feature disentanglement, enabling improved generalization to unseen classes, together with a two-stage training strategy that combines one-to-many matching and a diversity loss to promote representational diversity. Experimental results on the URBAN-SED and DESED datasets demonstrated that the proposed method performs competitively under the closed-world scenarios and surpasses existing approaches in open-world scenarios, thereby validating its effectiveness and robustness. Overall, this study lays the groundwork for advancing future research in open-world audio understanding, paving the way toward more adaptive, generalizable, and realistic sound event detection systems.

Future research will focus on extending OW-SED evaluation beyond the current benchmarks to larger-scale and more heterogeneous datasets that capture the variability of real-world acoustic environments, thereby providing a more rigorous assessment of scalability and robustness. Another promising direction is the integration of self-supervised pretraining and contrastive learning strategies, which can enrich the class-agnostic feature space and improve the transferability of learned representations to novel sound events with minimal supervision. In parallel, extending OW-SED to multimodal scenarios, particularly joint audio-visual event detection, could leverage complementary cues from vision to resolve challenges such as temporally overlapping or context-dependent acoustic events. Such multimodal extensions not only promise more robust and accurate detection but also improve the interpretability and applicability of OW-SED systems in complex real-world deployments.

\section*{Acknowledgments}
This research was funded by a project of Vietnam National University, Hanoi under Project No. QG.23.66.

\bibliographystyle{elsarticle-num}

\bibliography{references}

@INPROCEEDINGS{7096611,
  author={Mesaros, Annamaria and Heittola, Toni and Eronen, Antti and Virtanen, Tuomas},
  booktitle={2010 18th European Signal Processing Conference}, 
  title={Acoustic event detection in real life recordings}, 
  year={2010},
  volume={},
  number={},
  pages={1267-1271},
  doi={}}

@ARTICLE{7100934,
  author={Stowell, Dan and Giannoulis, Dimitrios and Benetos, Emmanouil and Lagrange, Mathieu and Plumbley, Mark D.},
  journal={IEEE Transactions on Multimedia}, 
  title={Detection and Classification of Acoustic Scenes and Events}, 
  year={2015},
  volume={17},
  number={10},
  pages={1733-1746},
  doi={10.1109/TMM.2015.2428998}}

@INPROCEEDINGS{7324337,
  author={Piczak, Karol J.},
  booktitle={2015 IEEE 25th International Workshop on Machine Learning for Signal Processing (MLSP)}, 
  title={Environmental sound classification with convolutional neural networks}, 
  year={2015},
  volume={},
  number={},
  pages={1-6},
  doi={10.1109/MLSP.2015.7324337}}

@article{crocco2016audio,
  title={Audio surveillance: A systematic review},
  author={Crocco, Marco and Cristani, Marco and Trucco, Andrea and Murino, Vittorio},
  journal={ACM Computing Surveys (CSUR)},
  volume={48},
  number={4},
  pages={1--46},
  year={2016},
  publisher={ACM New York, NY, USA}
}

@article{salamon2018sonyc,
  title={Sonyc: A system for the monitoring analysis and mitigation of urban noise pollution},
  author={Salamon, Justin and Bello, Juan and Silva, Claudio and Nov, Oded and DuBois, R and Arora, Anish and Mydlarz, Charlie and Doraiswamy, Harish},
  journal={Communications of the ACM},
  volume={5},
  year={2018}
}

@article{sedt,
  title={Sound event detection transformer: An event-based end-to-end model for sound event detection},
  author={Ye, Zhirong and Wang, Xiangdong and Liu, Hong and Qian, Yueliang and Tao, Rui and Yan, Long and Ouchi, Kazushige},
  journal={arXiv preprint arXiv:2110.02011},
  year={2021}
}

@inproceedings{phuong2013sound,
  title={Sound classification for event detection: Application into medical telemonitoring},
  author={Phuong, Nguyen Cong and Do Dat, Tran},
  booktitle={2013 International Conference on Computing, Management and Telecommunications (ComManTel)},
  pages={330--333},
  year={2013},
  organization={IEEE}
}

@inproceedings{cnn,
  title={Robust sound event recognition using convolutional neural networks},
  author={Zhang, Haomin and McLoughlin, Ian and Song, Yan},
  booktitle={2015 IEEE international conference on acoustics, speech and signal processing (ICASSP)},
  pages={559--563},
  year={2015},
  organization={IEEE}
}

@article{crnn,
  title={Convolutional recurrent neural networks for polyphonic sound event detection},
  author={Cak{\i}r, Emre and Parascandolo, Giambattista and Heittola, Toni and Huttunen, Heikki and Virtanen, Tuomas},
  journal={IEEE/ACM Transactions on Audio, Speech, and Language Processing},
  volume={25},
  number={6},
  pages={1291--1303},
  year={2017},
  publisher={IEEE}
}

@inproceedings{li2020sound,
  title={Sound event detection via dilated convolutional recurrent neural networks},
  author={Li, Yanxiong and Liu, Mingle and Drossos, Konstantinos and Virtanen, Tuomas},
  booktitle={ICASSP 2020-2020 IEEE International Conference on Acoustics, Speech and Signal Processing (ICASSP)},
  pages={286--290},
  year={2020},
  organization={IEEE}
}

@inproceedings{adavanne2017sound,
  title={Sound event detection using spatial features and convolutional recurrent neural network},
  author={Adavanne, Sharath and Pertil{\"a}, Pasi and Virtanen, Tuomas},
  booktitle={2017 IEEE international conference on acoustics, speech and signal processing (ICASSP)},
  pages={771--775},
  year={2017},
  organization={IEEE}
}

@inproceedings{hershey2017cnn,
  title={CNN architectures for large-scale audio classification},
  author={Hershey, Shawn and Chaudhuri, Sourish and Ellis, Daniel PW and Gemmeke, Jort F and Jansen, Aren and Moore, R Channing and Plakal, Manoj and Platt, Devin and Saurous, Rif A and Seybold, Bryan and others},
  booktitle={2017 ieee international conference on acoustics, speech and signal processing (icassp)},
  pages={131--135},
  year={2017},
  organization={IEEE}
}

@inproceedings{joseph2021towards,
  title={Towards open world object detection},
  author={Joseph, KJ and Khan, Salman and Khan, Fahad Shahbaz and Balasubramanian, Vineeth N},
  booktitle={Proceedings of the IEEE/CVF conference on computer vision and pattern recognition},
  pages={5830--5840},
  year={2021}
}

@inproceedings{prob,
  title={Prob: Probabilistic objectness for open world object detection},
  author={Zohar, Orr and Wang, Kuan-Chieh and Yeung, Serena},
  booktitle={Proceedings of the IEEE/CVF Conference on Computer Vision and Pattern Recognition},
  pages={11444--11453},
  year={2023}
}

@inproceedings{cat,
  title={Cat: Localization and identification cascade detection transformer for open-world object detection},
  author={Ma, Shuailei and Wang, Yuefeng and Wei, Ying and Fan, Jiaqi and Li, Thomas H and Liu, Hongli and Lv, Fanbing},
  booktitle={Proceedings of the IEEE/CVF Conference on Computer Vision and Pattern Recognition},
  pages={19681--19690},
  year={2023}
}

@inproceedings{ow-detr,
  title={Ow-detr: Open-world detection transformer},
  author={Gupta, Akshita and Narayan, Sanath and Joseph, KJ and Khan, Salman and Khan, Fahad Shahbaz and Shah, Mubarak},
  booktitle={Proceedings of the IEEE/CVF conference on computer vision and pattern recognition},
  pages={9235--9244},
  year={2022}
}

@article{d-detr,
  title={Deformable DETR: Deformable Transformers for End-to-End Object Detection},
  author={Zhu, Xizhou and Su, Weijie and Lu, Lewei and Li, Bin and Wang, Xiaogang and Dai, Jifeng},
  journal={arXiv preprint arXiv:2010.04159},
  year={2020}
}

@inproceedings{conformer,
  author = {Gulati, Anmol and Qin, James and Chiu, Chung-Cheng and Parmar, Niki and Zhang, Yu and Yu, Jiahui and Han, Wei and Wang, Shibo and Zhang, Zhengdong and Wu, Yonghui and Pang, Ruoming},
  booktitle = {INTERSPEECH},
  editor = {Meng, Helen and Xu, Bo and Zheng, Thomas Fang},
  ee = {https://doi.org/10.21437/Interspeech.2020-3015},
  pages = {5036-5040},
  publisher = {ISCA},
  title = {Conformer: Convolution-augmented Transformer for Speech Recognition.},
  year = 2020
}

@inproceedings{fdy-crnn,
  author={Hyeonuk Nam and Seong-Hu Kim and Byeong-Yun Ko and Yong-Hwa Park},
  title={{Frequency Dynamic Convolution: Frequency-Adaptive Pattern Recognition for Sound Event Detection}},
  year=2022,
  booktitle={Proc. Interspeech 2022},
  pages={2763--2767},
  doi={10.21437/Interspeech.2022-10127}
}

@inproceedings{ast-sed,
  title={Ast-sed: An effective sound event detection method based on audio spectrogram transformer},
  author={Li, Kang and Song, Yan and Dai, Li-Rong and McLoughlin, Ian and Fang, Xin and Liu, Lin},
  booktitle={ICASSP 2023-2023 IEEE International Conference on Acoustics, Speech and Signal Processing (ICASSP)},
  pages={1--5},
  year={2023},
  organization={IEEE}
}

@inproceedings{detr,
  title={End-to-end object detection with transformers},
  author={Carion, Nicolas and Massa, Francisco and Synnaeve, Gabriel and Usunier, Nicolas and Kirillov, Alexander and Zagoruyko, Sergey},
  booktitle={European conference on computer vision},
  pages={213--229},
  year={2020},
  organization={Springer}
}

@ARTICLE{conformer-sed,
  author={Barahona, Sara and de Benito-Gorrón, Diego and Toledano, Doroteo T. and Ramos, Daniel},
  journal={IEEE/ACM Transactions on Audio, Speech, and Language Processing}, 
  title={Enhancing Conformer-Based Sound Event Detection Using Frequency Dynamic Convolutions and BEATs Audio Embeddings}, 
  year={2024},
  volume={32},
  number={},
  pages={3896-3907},
  doi={10.1109/TASLP.2024.3444490}}

@article{openworlddetrtransformer,
  title={Open world DETR: Transformer based open world object detection},
  author={Dong, Na and Zhang, Yongqiang and Ding, Mingli and Lee, Gim Hee},
  journal={arXiv preprint arXiv:2212.02969},
  year={2022}
}

@article{ow-rcnn,
  title={Addressing the challenges of open-world object detection},
  author={Pershouse, David and Dayoub, Feras and Miller, Dimity and S{\"u}nderhauf, Niko},
  journal={arXiv preprint arXiv:2303.14930},
  year={2023}
}

@InProceedings{openmax,
title={Towards Open Set Deep Networks},
author={Bendale, Abhijit and Boult, Terrance},
booktitle={Computer Vision and Pattern Recognition (CVPR), 2016 IEEE Conference on},
year={2016},
organization={IEEE}
}

@inproceedings{doc,
    title = "{DOC}: Deep Open Classification of Text Documents",
    author = "Shu, Lei  and
      Xu, Hu  and
      Liu, Bing",
    editor = "Palmer, Martha  and
      Hwa, Rebecca  and
      Riedel, Sebastian",
    booktitle = "Proceedings of the 2017 Conference on Empirical Methods in Natural Language Processing",
    month = sep,
    year = "2017",
    address = "Copenhagen, Denmark",
    publisher = "Association for Computational Linguistics",
    url = "https://aclanthology.org/D17-1314/",
    doi = "10.18653/v1/D17-1314",
    pages = "2911--2916"
}

@inproceedings{medical,
  title={Open-world active learning for echocardiography view classification},
  author={Zamzmi, Ghada and Oguguo, Tochi and Rajaraman, Sivaramakrishnan and Antani, Sameer},
  booktitle={Medical Imaging 2022: Computer-Aided Diagnosis},
  volume={12033},
  pages={138--148},
  year={2022},
  organization={SPIE}
}

@INPROCEEDINGS{urban-sed,
  author={Salamon, Justin and MacConnell, Duncan and Cartwright, Mark and Li, Peter and Bello, Juan Pablo},
  booktitle={2017 IEEE Workshop on Applications of Signal Processing to Audio and Acoustics (WASPAA)}, 
  title={Scaper: A library for soundscape synthesis and augmentation}, 
  year={2017},
  volume={},
  number={},
  pages={344-348},
  doi={10.1109/WASPAA.2017.8170052}}

@inproceedings{ss-owdetr,
  title={Semi-supervised open-world object detection},
  author={Mullappilly, Sahal Shaji and Gehlot, Abhishek Singh and Anwer, Rao Muhammad and Khan, Fahad Shahbaz and Cholakkal, Hisham},
  booktitle={Proceedings of the AAAI Conference on Artificial Intelligence},
  volume={38},
  number={5},
  pages={4305--4314},
  year={2024}
}

@inproceedings{resnet,
  title={Deep residual learning for image recognition},
  author={He, Kaiming and Zhang, Xiangyu and Ren, Shaoqing and Sun, Jian},
  booktitle={Proceedings of the IEEE conference on computer vision and pattern recognition},
  pages={770--778},
  year={2016}
}

@inproceedings{chen2019endtoendaudioclassificationbased,
  title     = {An End-to-End Audio Classification System Based on Raw Waveforms and Mix-Training Strategy},
  author    = {Jiaxu Chen and Jing Hao and Kai Chen and Di Xie and Shicai Yang and Shiliang Pu},
  year      = {2019},
  booktitle = {Interspeech 2019},
  pages     = {3644--3648},
  doi       = {10.21437/Interspeech.2019-1579},
  issn      = {2958-1796},
}

@article{YIN2025109691,
title = {Multi-granularity acoustic information fusion for sound event detection},
journal = {Signal Processing},
volume = {227},
pages = {109691},
year = {2025},
issn = {0165-1684},
doi = {https://doi.org/10.1016/j.sigpro.2024.109691},
author = {Han Yin and Jianfeng Chen and Jisheng Bai and Mou Wang and Susanto Rahardja and Dongyuan Shi and Woon-seng Gan}
}

@article{ZHENG2026110218,
title = {STWWgram-ODCBAM: Multimodal feature fusion and dynamic attention mechanism for anomalous sound detection},
journal = {Signal Processing},
volume = {239},
pages = {110218},
year = {2026},
issn = {0165-1684},
doi = {https://doi.org/10.1016/j.sigpro.2025.110218},
url = {https://www.sciencedirect.com/science/article/pii/S0165168425003329},
author = {Libin Zheng and Dongsheng Liu and Tong Wu and Yahui Chen}
}

@article{osse,
author = {You, Jie and Wu, Wenqin and Lee, Joonwhoan},
year = {2024},
month = {01},
pages = {},
title = {Open set classification of sound event},
volume = {14},
journal = {Scientific Reports},
doi = {10.1038/s41598-023-50639-7}
}

@misc{detectanysound,
      title={Detect Any Sound: Open-Vocabulary Sound Event Detection with Multi-Modal Queries}, 
      author={Pengfei Cai and Yan Song and Qing Gu and Nan Jiang and Haoyu Song and Ian McLoughlin},
      year={2025},
      eprint={2507.16343},
      archivePrefix={arXiv},
      primaryClass={cs.SD},
      url={https://arxiv.org/abs/2507.16343}, 
}

@misc{flexsed,
      title={FlexSED: Towards Open-Vocabulary Sound Event Detection}, 
      author={Jiarui Hai and Helin Wang and Weizhe Guo and Mounya Elhilali},
      year={2025},
      eprint={2509.18606},
      archivePrefix={arXiv},
      primaryClass={eess.AS},
      url={https://arxiv.org/abs/2509.18606}, 
}

@misc{ucil,
      title={UCIL: An Unsupervised Class Incremental Learning Approach for Sound Event Detection}, 
      author={Yang Xiao and Rohan Kumar Das},
      year={2025},
      eprint={2407.03657},
      archivePrefix={arXiv},
      primaryClass={eess.AS},
      url={https://arxiv.org/abs/2407.03657}, 
}

@misc{pandey2024classincrementallearningsoundevent,
      title={Class-Incremental Learning for Sound Event Localization and Detection}, 
      author={Ruchi Pandey and Manjunath Mulimani and Archontis Politis and Annamaria Mesaros},
      year={2024},
      eprint={2411.12830},
      archivePrefix={arXiv},
      primaryClass={eess.AS},
      url={https://arxiv.org/abs/2411.12830}, 
}

@inproceedings{desed,
  title={Sound event detection in domestic environments with weakly labeled data and soundscape synthesis},
  author={Turpault, Nicolas and Serizel, Romain and Shah, Ankit Parag and Salamon, Justin},
  booktitle={Workshop on Detection and Classification of Acoustic Scenes and Events},
  year={2019}
}

\end{document}